\documentclass{aastex63}

\usepackage{amssymb}
\usepackage{amsmath}
\usepackage{xcolor}
\usepackage{makecell}
\usepackage{booktabs}
\usepackage{natbib}
\usepackage{lineno}
\submitjournal{ApJ}
\shorttitle{Chen et al.}
\graphicspath{{./}{figures/}}
\begin{document}
\title{FRBs Lensed by Point Masses II. The multi-peaked FRBs from the point view of microlensing}

\author{Xuechun Chen}
\affil{Purple Mountain Observatory, Chinese Academy of Sciences, Nanjing, Jiangsu, 210023, China}
\affil{School of Astronomy and Space Science, University of Science and Technology of China, Hefei, Anhui, 230026, China}
\author{Yiping Shu}
\affil{Max-Planck-Institut f\"{u}r Astrophysik, Karl-Schwarzschild-Str. 1, 85748 Garching, Germany}
\affil{Ruhr University Bochum, Faculty of Physics and Astronomy, Astronomical Institute (AIRUB), German Centre for Cosmological Lensing, 44780 Bochum, Germany}
\author{Guoliang Li}
\affil{Purple Mountain Observatory, Chinese Academy of Sciences, Nanjing, Jiangsu, 210023, China}
\email{guoliang@pmo.ac.cn}
\author{Wenwen Zheng}
\affil{Purple Mountain Observatory, Chinese Academy of Sciences, Nanjing, Jiangsu, 210023, China}
\affil{School of Astronomy and Space Science, University of Science and Technology of China, Hefei, Anhui, 230026, China}
\correspondingauthor{Guoliang Li}
\begin{abstract}
The microlensing effect has developed into a powerful technique for a diverse range of applications including exoplanet discoveries, structure of the Milky Way, constraints on MAssive Compact Halo Objects, and measurements of the size and profile of quasar accretion discs. In this paper, we consider a special type of microlensing events where the sources are fast radio bursts with $\sim$milliseconds (ms) durations for which the relative motion between the lens and source is negligible. In this scenario, it is possible to temporally resolve the individual microimages. As a result, a method beyond the inverse ray shooting (IRS) method, which only evaluates the total magnification of all microimages, is needed. We therefore implement an algorithm for identifying individual microimages and computing their magnifications and relative time delays. We validate our algorithm by comparing to analytical predictions for a single microlens case and find excellent agreement. We show that the superposition of pulses from individual microimages produces a light curve that appears as multi-peaked FRBs. The relative time delays between pulses can reach 0.1--1 ms for stellar-mass lenses and hence can already be resolved temporally by current facilities. Although not yet discovered, microlensing of FRBs will become regular events and surpass the number of quasar microlensing events in the near future when $10^{4-5}$ FRBs are expected to be discovered on a daily basis. Our algorithm provides a way of generating the microlensing light curve that can be used for constraining stellar mass distribution in distant galaxies.
\end{abstract}

\section{Introduction} \label{sec:intro}

Following the discovery of the first strongly-lensed quasars \citep{1979Natur.279..381W}, \citet{Chang1979, Chang1984} pointed out that stars in the lensing galaxy can produce additional lensing effects when the size of the background source (such as quasars) is comparable to or smaller than the Einstein radii of the stars, in which the macroimage of quasars can be splitted into multiple sub-images. \citet{Paczynski86a} used the terms ``microlensing'' and ``microimages'' to describe the lensing effect from stars and further proposed an idea of searching for MAssive Compact Halo Objects (MACHOs) in the Milky Way using the microlensing effect \citep{Paczynski86b}. Since the first detection of a microlensing event \citep{1989AJ.....98.1989I}, microlensing has played an outstanding role in a wide range of fields, such as discovering exoplanets and binary stars \citep[e.g.,][]{1991ApJ...374L..37M,1992ApJ...396..104G}, inferring the structure of the Milky Way \citep[e.g.,][]{Udalski2003, Hamadache2006}, constraining the fraction of dark matter in the form of MACHOs \citep[e.g.,][]{2017ApJ...836L..18M,2018ASPC..514...79S}, determining the ratio of stellar to dark matter \citep[e.g.,][]{2002ApJ...580..685S,2004IAUS..220..103S}, measuring the size and profile of quasar accretion discs \citep[e.g.,][]{2016ApJ...830..149F,2018ApJ...869..132F}, etc. 
 
As the separations among microimages are on the order of micro-arcsecond ($\mu$as) and therefore unresolved in most cases, microlensing-related studies so far have relied on analysing the light curve that is the change of the total brightness of all lensed microimages as the lens-source alignment changes due to relative motions. In this paper, we will discuss a special type of microlensing events where the background source is a transient with such a short duration that the microimages can be resolved temporally. As will be shown later, the typical time delays between microimages are $\sim$0.1--1 millisecond (ms). We therefore take fast radio bursts (FRBs), the durations of which are on the order of ms, as an example.  

FRBs are bright radio transient signals with durations on the order of milliseconds \citep{2019A&ARv..27....4P, Xiao2021}. They are detected at frequencies from hundreds of MHz to several GHz, and the arrival time of signals in different frequencies is usually different due to the dispersion effect caused by cold plasma \citep[e.g.,][]{Lorimer2007}. The lag in arrival time is proportional to the dispersion measure (DM), which is defined as the integration of the column density of free electrons along the line of sight (LOS). By measuring the DM from time lag and comparing it to the expected contribution of the Milky Way, it is clear that most FRBs are extragalactic. In fact, several FRB sources have been located to their host galaxies. For example, FRB121102 was precisely localized to reside in a dwarf star forming galaxy at $z_{s}=0.1927$ \citep{Chatterjee2017, Tendulkar2017} and FRB180916 was located to reside in a nearby spiral galaxy at $z_{s}=0.0337$ \citep{2020Natur.577..190M}. The formation mechanism of the bursts is still an open question since the first discovery by \citet{Lorimer2007}. The properties of FRB sources also seem heterogeneous. Among the discovered FRB samples, the majority show a single burst while some were found to have additional bursts \citep{2016Natur.531..202S,2019Natur.566..235C,2019ApJ...885L..24C,2020ApJ...891L...6F}, which are usually referred to as non-repeating and repeating FRBs. 

Given the unique characteristics such as point like, transient, high event rate and extragalactic origin, there have been a lot of discussions in the literature on gravitational lensing effects on FRBs. For instance, \citet{Munoz2016} discussed an idea of constraining the amount of dark matter in the form of MACHO using strongly-lensed FRBs. \citet{DaiLu2017} proposed to measure the motion of the emission regions of strongly-lensed repeating FRBs from the variations of time delays, which could place constraints on the physical nature of the emissions. \citet{Wagner2018} suggested to reconstruct the mass distribution of the lens with higher accuracy using strongly-lensed repeating FRBs. \citet{Li2017} proposed to constrain cosmological parameters using strongly-lensed repeating FRBs. 
\citet{Lewis2020} suggested that the behaviours of microimages of FRBs can be used to reveal the parities of the macroimages, which provides additional constraints on the macro lens model. 
In our previous work \citep{2021ApJ...912..134C}, we showed that lens mass constraints can be obtained from the observed time delay and flux ratio of a FRB strongly lensed by a point-mass object. 


In this work, we continue to explore the microlensing effect of FRBs. Since the duration of FRBs is on the order of millisecond, comparable to typical time delays (0.1--1ms) in stellar-mass microlensing, it is possible to temporally resolve the individual microimages. As a result, information of individual microimages (i.e. positions, magnifications, and time delays) is desired. A method beyond the inverse ray shooting (IRS) method, which only evaluates the total magnification of all microimages, is needed. The paper is organized as follows. Section \ref{sec:lensTheory} describes the basic theory of microlensing. Section \ref{sec:Sim} introduces an algorithm we implement for finding microimages. Validations of the algorithm are presented in Section \ref{sec:test}. Section \ref{sec:discussion} provides some discussions on microlensing of FRBs and a summary is given in Section \ref{sec:conclusion}. Throughout the paper, we assume a fiducial cosmology of $\Omega_m = 0.272, \Omega_{\lambda} = 0.728, H_0=70.4$ km s$^{-1}$ Mpc$^{-1}$\citep{Komatsu2011}.

\section{The theory of microlensing}
\label{sec:lensTheory}
In this work, we focus on the microlensing effect of stars embedded in a distant lens object (e.g. a galaxy). As a result, we adopt the thin lens approximation in which the lensing effect is determined by the projected effective lensing potential. Because the angular scales of the microlensing effect considered here are on the order of microarcsecond ($\mu$as), the localised effective lensing potential can be used. The lens surface mass distribution, which determines the effective lensing potential, is assumed to consist of two components: the smoothly distributed matter (e.g. dark matter) and the compact objects (e.g. stars). It is usually convenient to work with convergence that is the surface mass density normalised by the critical density $\Sigma_{0}=\frac{1}{\pi}\frac{c^{2}}{4G}\frac{D_{s}}{D_{l}D_{ls}}$, where $D_{l}$, $D_{s}$, and $D_{ls}$ are the angular diameter distances from the observer to the lens, from the observer to the source, and from lens to the source respectively. The total convergence can therefore be written as $\kappa=\kappa_{s}+\kappa_{\ast}$ where $\kappa_{s}$ and $\kappa_{\ast}$ denote the convergences from smoothly distributed matter and stars respectively. 

We start with a simple case where a single star with mass $M$ is included. In this case, the total convergence $\kappa \approx \kappa_{s}$. The localised effective lensing potential can be written as
\begin{eqnarray}
\label{eq:phi-single-star}
\psi \left ( \vec{\theta} \right )=\frac{\theta _{E}^{2}}{2}\ln \left | \vec{\theta}\right |^{2}+\frac{\kappa _{s}}{2}\left ( \theta_{1}^{2}+\theta_{2}^{2}\right )-\frac{\gamma}{2}\left ( \theta_{1}^{2}-\theta_{2}^{2}\right ).
\end{eqnarray}
Here $\theta_{E}$ is the Einstein radius of the star defined as 
\begin{eqnarray*}
\label{eq:thetaE}
\theta _{E}=\sqrt{\frac{4GM}{c^{2}}\, \frac{D_{ls}}{D_{l}D_{s}}}
\end{eqnarray*}
and $\gamma$ is the shear from the lensing galaxy as a whole. We note that $\kappa_{s}$ is approximated as a constant as its relative variation over the region under consideration is on the level of $\sim 10^{-3}$. The lens equation can be written as
\begin{eqnarray}
\label{eq:single-star}
\vec{\beta}=\begin{pmatrix}
1-\kappa_{s}+\gamma & 0\\ 
 0&1-\kappa_{s}-\gamma 
\end{pmatrix}\vec{\theta}-\theta_{E}^2\frac{\vec{\theta}}{\left | \vec{\theta} \right |^2},
\end{eqnarray}
Here $\vec{\beta}$ is the source position in the source plane and $\vec{\theta}$ is the image position in the deflector plane. The second term of the right side in Equation (\ref{eq:single-star}) is the contribution from the single star. By introducing $\lambda_{r}=1-\kappa+\gamma$, $\lambda_{t}=1-\kappa-\gamma$, $\vec{y}=\vec{\beta}/\theta_{E}$, and $\vec{x}=\vec{\theta}/\theta_{E}$, Equation (\ref{eq:single-star}) reads in a two-dimensional form as:
\begin{eqnarray}
\label{eq:scale-eq-1}
y_{1}&=&\lambda _{r}x_{1}-\frac{x_{1}}{x_{1}^2+x_{2}^2},\\
\label{eq:scale-eq-2}
y_{2}&=&\lambda _{t}x_{2}-\frac{x_{2}}{x_{1}^2+x_{2}^2}.
\end{eqnarray}
where subscripts 1 and 2 denote the two dimensions.\\

In this scenario, two or four lensed images (referred to microimages hereafter) can be produced. In particular, for source positions along the coordinate axes (i.e. $y_{1}=0$ or $y_{2}=0$), Equations (\ref{eq:scale-eq-1}) and (\ref{eq:scale-eq-2}) can be solved analytically. For $y_{2}=0$, the lensed images are located at
\begin{eqnarray}
\label{eq:ana-root1}
x_{1}&=&\frac{y_{1}\pm \sqrt{y_{1}^2+4\lambda _{r}}}{2\lambda _{r}},\;x2=0\\
\label{eq:ana-root11}
x_{1}&=&\frac{y_{1}}{\lambda_{r}-\lambda_{t}},\;x_{2}=\pm \sqrt{\frac{(\lambda_{r}-\lambda_{t})^2-\lambda_{t}y_{1}^2}{\lambda_{t}(\lambda_{r}-\lambda_{t})^2}} \text{ (if $\lvert y_1 \rvert \leq \sqrt{\frac{(\lambda_{r}-\lambda_{t})^2}{\lambda_{t}}}$)};
\end{eqnarray}
and for $y_{1}=0$, the lensed images are located at
\begin{eqnarray}
\label{eq:ana-root2}
x_{1}&=&0,\;x_{2}=\frac{y_{2}\pm \sqrt{y_{2}^2+4\lambda_{t}}}{2\lambda_{t}},\\
\label{eq:ana-root22}
x_{1}&=&\pm \sqrt{\frac{(\lambda_{t}-\lambda_{r})^2-\lambda_{r}y_{2}^2}{\lambda_{r}(\lambda_{t}-\lambda_{r})^2}},\;x_{2}=\frac{y_{2}}{\lambda_{t}-\lambda_{r}}  \text{ (if $\lvert y_2 \rvert \leq \sqrt{\frac{(\lambda_{t}-\lambda_{r})^2}{\lambda_{r}}}$)}.
\end{eqnarray}
Obviously, the intersections of the caustic with the axes are given by $y_{1}=\pm \sqrt{\frac{(\lambda_{r}-\lambda_{t})^2}{\lambda_{t}}}$ and $y_{2}=\pm \sqrt{\frac{(\lambda_{t}-\lambda_{r})^2}{\lambda_{r}}}$.\\

Now we consider a more general situation where there are $N$ stars. For simplicity, we assume all the $N$ stars have the same mass $M$. The localised effective lensing potential becomes
\begin{eqnarray}
\label{eq:phi}
\psi \left ( \vec{\theta }\right )=\frac{\theta_{E}^{2}}{2}\sum_{i}^{N}\ln\left | \vec{\theta}-\vec{\theta}^{(i)}\right |^{2}+\frac{\kappa_{s}}{2}\left ( \theta_{1}^{2}+\theta_{2}^{2}\right )-\frac{\gamma}{2}\left ( \theta_{1}^{2}-\theta_{2}^{2}\right ),
\end{eqnarray}
where $\theta_{E}$ is the Einstein radius determined by mass $M$ as expressed above. $\vec{\theta}^{(i)}$ is the position of the $i^{th}$ star (also referred to as a microlens). Again, $\kappa_s$ is approximated as a constant. 
The lens equation can be written as:
\begin{eqnarray}
\label{eq:many-stars}
\vec{\beta}=\begin{pmatrix}
1-\kappa_{s}+\gamma & 0 \\ 
0 & 1-\kappa_{s}-\gamma
\end{pmatrix}\vec{\theta}-\theta_{E}^2\sum_{i}^{N}\frac{\vec{\theta}-\vec{\theta}^{(i)}}{\left | \vec{\theta}-\vec{\theta}^{(i)}\right |^2}.
\end{eqnarray}
$\kappa$, $\gamma$, $\kappa_{\ast}$ and the mass function of micro-lenses determine the properties of microlensing \citep{2012ApJ...744...90B}. Similarly, we rewrite Equation (\ref{eq:many-stars}) in the two-dimensional form with $\vec{y}=\vec{\beta}/\theta_{E}$, $\vec{x}=\vec{\theta}/\theta_{E}$
\begin{equation}
\label{eq:scale-many-stars-1}
y_{1}=x_{1}-\alpha_{1}, \; \alpha_{1}=\sum_{i}^{N}\frac{x_{1}-x^{(i)}_{1}}{\left |\vec{x}-\vec{x}^{(i)} \right |^{2}}+\kappa_{s}x_{1}-\gamma x_{1},
\end{equation}
\begin{equation}
\label{eq:scale-many-stars-2}
y_{2}=x_{2}-\alpha_{2}, \; \alpha_{2}=\sum_{i}^{N}\frac{x_{2}-x^{(i)}_{2}}{\left |\vec{x}-\vec{x}^{(i)} \right |^{2}}+\kappa_{s}x_{2}+\gamma x_{2}.
\end{equation}
\\
For an image at position $\vec{x} =\left ( x_{1}, \; x_{2}\right )$ in the deflector plane, the magnification is given by
\begin{eqnarray}
\label{eq:mu}
\mu ^{-1}=\left ( 1-\frac{\partial\alpha_{1}}{\partial x_{1}}\right )\left ( 1-\frac{\partial\alpha_{2}}{\partial x_{2}}\right )-\frac{\partial\alpha_{1}}{\partial x_{2}}\frac{\partial\alpha_{2}}{\partial x_{1}}.
\end{eqnarray}
The time-delay function is
\begin{eqnarray}
\label{eq:delta_t_function}
t=\frac{D_{l}D_{s}}{cD_{ls}}\left ( 1+z_{l}\right )\left [ \frac{1}{2}\left ( \vec{\theta}-\vec{\beta}\right )^{2}-\psi \left ( \vec{\theta}\right )\right ].
\end{eqnarray}
For an image at $\vec{x}=\left ( x_{1}, \; x_{2}\right )$, the time delay relative to the unlensed case is
\begin{align}
\label{eq:scaled_delta_t}
\Delta t &= \frac{2GM\left ( 1+z_{l}\right )}{c^{3}}\left [ \left ( \vec{x}-\vec{y}\right )^{2}-\sum_{i}^{N} \ln\left ( \vec{x}^{(i)}-\vec{x}\right )^{2}-\kappa_{s}\left ( {x_{1}}^{2}+{x_{2}}^{2} \right )+\gamma\left ({x_{1}}^{2}-{x_{2}}^{2} \right )-2 N \ln \theta_{E}\right ] \nonumber \\
&= (9.85 \mu \text{s}) (1+z_l) (\frac{M}{M_{\odot}})\left [ \left ( \vec{x}-\vec{y}\right )^{2}-\sum_{i}^{N} \ln\left ( \vec{x}^{(i)}-\vec{x}\right )^{2}-\kappa_{s}\left ( {x_{1}}^{2}+{x_{2}}^{2} \right )+\gamma\left ({x_{1}}^{2}-{x_{2}}^{2} \right )-2 N \ln \theta_{E}\right ].
\end{align}
We note that the magnification $\mu$ and the absolute time delay relative to the unlensed case are not direct observables because the intrinsic luminosity and position of the background source are usually unknown. Nevertheless, the relative time delay between microimages is an observable.\\

\begin{figure}
\centerline{\scalebox{1.0}
{\includegraphics[width=0.5\textwidth]{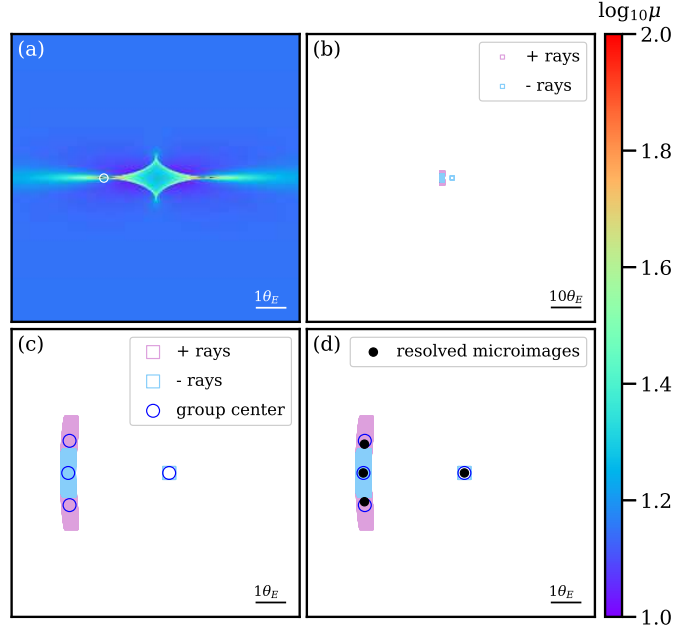}}}
\caption{\label{fig:fig1} Illustration of the processes of finding microimages in our method. The lensing field used is a $1M_{\odot}$ star $+$ convergence $\kappa=0.6$ and shear $\gamma=0.3$. Panel (a)  shows the magnification map on the logarithmic scale (i.e. $\log_{10} \mu$) in a square region of the source plane with side length L$=10\theta_{E}$. The open white circle shows the position of the source. Panel (b) shows the ray-emitting region in the deflector plane. The side of this panel is S=$100\theta_{E}$. The open purple and blue squares correspond to pixel positions of light rays with positive and negative parity that fall within $r_{s}$ of the target source in panel (a). Panel (c) is an enlarged view of the deflector plane. The blue circles denote the centroids of the three clusters identified by the friend-of-friend algorithm. The black dots in panel (d) show the positions of final resolved microimages after the cleaning and merging processes.}
\end{figure}

In this scenario, the number of microimages is not limited to two or four, and the lens equation generally can not be solved analytically. Nevertheless, microlensing-related studies so far do not attempt to resolve the individual microimages because the required angular resolution is beyond the capabilities of all current facilities. Hence, those studies primarily make use of the total magnification (i.e. summed over microimages) and its variation with respect to the position change between the lens and the source. As a result, the inverse ray shooting (IRS) method has been extensively used to generate the magnification map \citep[e.g.,][]{Kayser1986, 1987A&A...171...49S, 1999JCoAM.109..353W}. In the IRS method, numerous light rays are emitted backwards from the observer to the deflector plane and eventually land in the source plane as determined by the deflection angle. Usually the deflector plane is divided into a regular pixel grid where each pixel corresponds to one light ray. To infer the magnification in any given pixel, the location of that light ray in the source plane is identified (i.e. source position) and light rays within a certain distance to that source position are selected. The ratio of the areas covered by those light rays in the deflector plane to that in the source plane is taken as the total magnification in that pixel in the source plane. It is clear that the so-generated magnification map is only an approximation, and the accuracy depends on the resolution in the deflector plane.\\ 

In this work, we are interested in the microlensing effect of a special type of sources --- transient objects --- that have such short durations (for example, FRBs) that the microimages can be resolved temporally. In this scenario, the positions of each individual microimages need to be obtained first, after which their magnification and time delay can be calculated straightforwardly from Equations (\ref{eq:mu}) and (\ref{eq:delta_t_function}). As a result, an approach different from the current IRS-based approaches, which can not provide microimage positions, is needed. We have therefore implemented a numerical method for finding the microimages, which can not be solved analyticall in general.   

\section{Methodology}
\label{sec:Sim}

The first step in our method is to also shoot light rays backwards from the observer to the source plane via the deflector plane. We start with an interested square region in the source plane of side length L (i.e., from $-\frac{\rm L}{2}$ to $\frac{\rm L}{2}$) that is chosen to be equal to 10$\theta_{\rm E}$. In addition, we add a buffer of length 10$\sqrt{\kappa_{\ast}}$ around this square region when calculating the required ray-emitting region in the deflector plane because the gravitational scattering of randomly distributed stars in the deflector plane can have an effect on the edge of the source plane \citep{1986ApJ...306....2K}. Under the effect of convergence $\kappa$ and shear $\gamma$, a square of length unity in the deflector plane will be mapped to a rectangle in the source plane with length and width of $1-\kappa+\gamma$ and $1-\kappa-\gamma$. We therefore set the ray-emitting region in the deflector plane as a square with side length S (i.e., from $-\frac{\rm S}{2}$ to $\frac{\rm S}{2}$), where S is $\rm Max\left [ \left |\frac{\rm L+20\sqrt{\kappa_{\ast}}}{1-\kappa+\gamma} \right |,\;\left |\frac{\rm L+20\sqrt{\kappa_{\ast}}}{1-\kappa-\gamma} \right |\right ]$. Microlenses are randomly distributed in the circumcircle of that square ray-emitting region in the deflector plane and the mass of each microlens is set to be 1 $M_{\odot}$. Using a circular microlens field is a standard procedure in microlensing studies \citep[e.g.][]{Kayser1986, 1990PhDT.......180W}. The number of microlenses---$N$ is therefore determined from $\kappa_{\ast}$ and the area of the circumcircle as $N=\frac{M_{\rm total}}{M}=\frac{\pi S^{2}D_{l}^{2}\kappa_{\ast}\Sigma_{0}}{2M}$. Hence, the microlensing effect is completely described by the local properties: $\kappa$, $\gamma$, and $\kappa_{\ast}$.

The square region in the deflector plane is divided into $\rm N_{\rm r}$ $\times$ $\rm N_{\rm r}$ pixels with each pixel corresponding to one light ray from the observer. The resolution of the deflector plane is therefore $r_{l}$=S/$\rm N_{\rm r}$. The value of $r_{l}$ should be as small as possible with the affordability of computing resource and time consuming. Rays are mapped through the field of micro-lenses to the source plane, and a mapping between the positions of light rays in the deflector plane and the source plane is obtained. We use $\left \{\vec{x}\right \}$ sample and $\left \{\vec{y}\right \}$ sample to denote the positions of rays in the deflector plane and their corresponding positions in the source plane.


Now for a given position $\vec{y}_{o}$ in the source plane (i.e. panel (a) of Figure~\ref{fig:fig1}), we identify rays from the $\left \{\vec{y}\right \}$ sample whose angular distance to $\vec{y}_{o}$ are within $r_{s}$ and call them $\left \{\vec{y}_{n}\right \}$ sample. 
$r_{s}$ is related to $r_{l}$ as
\begin{eqnarray*}
r_{s}=\sqrt{\frac{n_{s}}{\pi \mu }}r_{l}.
\end{eqnarray*}
where $n_{s}$ is the expected number of rays in the $\left \{\vec{y}_{n}\right \}$ sample and $\mu$ is the total magnification. Obviously, the value of $r_{s}$ should be large enough that at least one light ray can be identified. In this work, we choose $r_{s}=5r_{l}$, which allows us to identify at least one light ray from microimages that have magnifications larger than $\mu_{\rm limit} =0.01273$. A larger $r_s$ can be used as long as the computing time permits. We then trace back to the deflector plane and group the light rays that correspond to the $\left \{\vec{y}_{n}\right \}$ sample as the $\left \{\vec{x}_{n}\right \}$ sample (i.e. all the open squares in panel (b) of Figure~\ref{fig:fig1}). We divide the $\left \{\vec{x}_{n}\right \}$ sample into $\left \{\vec{x}_{n}^{+}\right \}$ sample and $\left \{\vec{x}_{n}^{-}\right \}$ sample according to their parity, which is necessary for resolving close pairs of microimage and its counter image (i.e. magenta and blue squares in panel (b) of Figure~\ref{fig:fig1}). The $\left \{\vec{x}_{n}^{+}\right \}$ sample and $\left \{\vec{x}_{n}^{-}\right \}$ sample are grouped into $n_{p}^{+}$ positive-parity clusters and $n_{p}^{-}$ negative-parity clusters using the friend-of-friend algorithm with a clustering criterion of $r_{p}$, which is chosen to be $\sim 1.5r_{l}$. The centroids of each cluster are recorded as [${\vec{x}_{c}}^{1}$, ${\vec{x}_{c}}^{2}$,..., ${\vec{x}_{c}}^{n_p}$] where $n_p=n_{p}^{+} + n_{p}^{-}$ (i.e. panel (c) of Figure~\ref{fig:fig1}).
Then we perform a root finding for Equation (\ref{eq:scale-many-stars-1}) and Equation (\ref{eq:scale-many-stars-2}) using the two-dimensional Newton iteration method with the centroids of each cluster obtained above as the starting points. The convergence criterion is set to $10^{-6}$. As a result, $n_{p}$ numerical solutions [${\vec{x}_{\rm rt1}}^{1}$, ..., ${\vec{x}_{\rm rt1}}^{n_{p}}$] are obtained for the $n_{p}$ clusters. For each position in the $\left \{\vec{x}_{\rm rt1}\right \}$ sample, we further find its mirror position with respect to the corresponding cluster centroid $\left \{\vec{x}_{c}\right \}$ and use it as the new starting point to perform the root finding again, which results in another $n_{p}$ numerical solutions [${\vec{x}_{\rm rt2}}^{1}$, ..., ${\vec{x}_{\rm rt2}}^{n_{p}}$]. The purpose of this second iteration is also to avoid mistaking a pair of closely located microimage and its counter image as one image.

To clean up roots from the above step, the $\left \{\vec{x}_{\rm rt1}\right \}$ sample and $\left \{\vec{x}_{\rm rt2}\right \}$ sample are combined together and divided into $\left \{\vec{x}_{\rm rt+}\right \}$ group: [$\vec{x}_{\rm rt+}^1$, ..., $\vec{x}_{\rm rt+}^{n+}$] and $\left \{\vec{x}_{\rm rt-}\right \}$ group: [$\vec{x}_{\rm rt-}^1$, ..., $\vec{x}_{\rm rt-}^{n-}$] according to their parities. Then the friend-of-friend algorithm is performed again to the $\left \{\vec{x}_{\rm rt+}\right \}$ sample and $\left \{\vec{x}_{\rm rt-}\right \}$ sample separately, where two roots $\vec{x}^{i}$ and $\vec{x}^{j}$ that have separation smaller than $(\rm Max\left [ \mu^{i},\; \sqrt{\mu^{i}}\right ]+\rm Max\left [ \mu^{j},\; \sqrt{\mu^{j}}\right ])$ are merged into one root, and the final root list $\left \{\vec{x}_{\rm Img}\right \}$ (including $\left \{\vec{x}_{\rm Img+}\right \}$ and $\left \{\vec{x}_{\rm Img-}\right \}$) is obtained (i.e. panel (d) of Figure~\ref{fig:fig1}). The magnification and time delay for each of the microimages can be computed directly from Equations (\ref{eq:mu}) and (\ref{eq:delta_t_function}).

\section{Validations}

\label{sec:test}
\subsection{Single star case}
\label{subsec:single-star}
The first test we did is for the case of a single star embedded in a field of convergence $\kappa=0.6$ and shear $\gamma=0.3$. 
As discussed in Section~\ref{sec:lensTheory}, the lens equation can be solved analytically in this case when the source is located on the coordinate axes. We therefore select 1000 uniformly distributed source positions in the range of $\left ( -5\theta_{E},\;5\theta_{E}\right )$ along the $y_{1}$-axis and $y_{2}$-axis (2000 sources in total). Analytically, we know that the 1000 source positions along the $y_1$ axis correspond to 621 two-image cases and 379 four-image cases, while the 1000 source positions along the $y_2$ axis correspond to 857 two-image cases and 143 four-image cases. We then calculate for each source position the number, positions, and magnifications of all its microimages found by our method, and compare the results with analytical solutions.

\begin{figure}
\centerline{\scalebox{1.0}
{\includegraphics[width=0.5\textwidth]{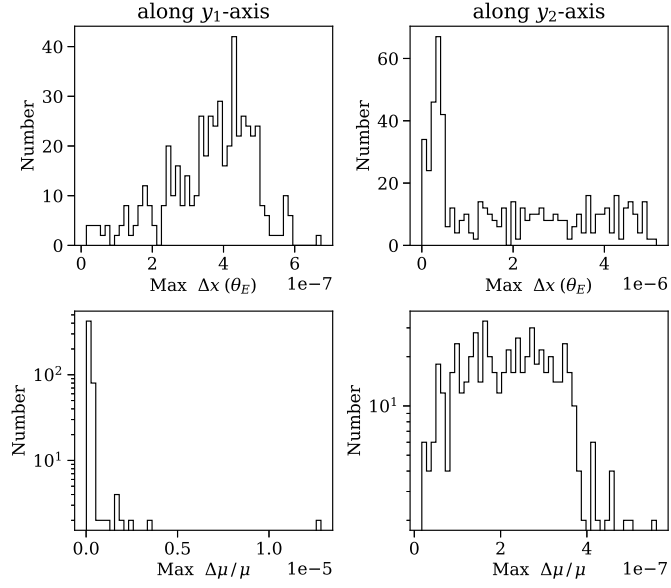}}}
\caption{\label{fig:fig2} Distributions of maximum differences in position and maximum relative differences in magnification of microimages given by our method and the analytical approach for the first test in Section~\ref{sec:test}. }
\end{figure}

\begin{deluxetable}{ccccccccccccccccccccc}
\tablewidth{0pt}
\tablecaption{Incidence rates of missing microimages. (Only microimages with magnification above $\mu_{\rm limit}$ are shown.).\label{tab:table1}}
\tablehead{ 
 \colhead{} & \colhead{2-image region} & \colhead{4-image region} & \colhead{2-image region} & \colhead{4-image region}\\
 \colhead{$r_{l}$} & \colhead{along $y_{1}$-axis} & \colhead{along $y_{1}$-axis} & \colhead{along $y_{2}$-axis} & \colhead{along $y_{2}$-axis}}
\startdata 
0.02 & 0/158 & 18/379 & 0/456 & 18/143  \\ \hline
0.01 & 0/158 & \phantom{1}8/379 & 0/456 & 10/143  \\ \hline
0.005 & 0/158 & \phantom{1}2/379 & 0/456 & \phantom{1}4/143  \\
\enddata
\end{deluxetable}

We first examine the incidence of missing microimages. It is realised that there are two types of missing microimages in our method. The first is related to the finite search radius $r_s$. Microimages with magnifications below a certain limit will simply be missed because their corresponding light rays fall outside the search radius $r_s$ (Type I). For the fiducial choice of $r_s=5 r_l$, this lower limit in magnification is $\mu_{\rm limit} =0.01273$. The other type of missing microimages is due to the failure in the root finding process (Type II), which we think is more relevant. Among the 2000 test cases, microimages in all the four-image cases have magnifications above $\mu_{\rm limit}$. For the 621 two-image cases along $y_1$ axis, 158 cases have both microimages with magnifications $\geqslant \mu_{\rm limit}$ and the remaining 463 cases have one microimage with magnification $< \mu_{\rm limit}$. For the 857 two-image cases along $y_2$ axis, 456 cases have both microimages with magnifications $\geqslant \mu_{\rm limit}$ and the remaining 401 cases have one microimage with magnification $< \mu_{\rm limit}$. 

We consider three choices of $r_l$, i.e. 0.02$\theta_{E}$, 0.01$\theta_{E}$, and 0.005$\theta_{E}$. The results are summarised in Table \ref{tab:table1}. We find that there is no Type II miss for any of the two-image test cases, regardless of the $r_l$ value. Missing of microimages (Type II) occurs in 6.9\% of all 522 four-image cases when $r_l = 0.02 \theta_{E}$. These numbers drops to 1.1\% when $r_l = 0.005 \theta_{E}$.
For situations that all microimages are successfully identified, we examine the maximum differences in image position and maximum relative differences in magnification obtained by our method (with $r_l=0.02 \theta_{\rm E}$) and the analytical approach in Figure \ref{fig:fig2}. It is clear that our method can reproduce the positions of microimages with an accuracy of better than 1--5$\times 10^{-6} \theta_{\rm E}$, which is essentially determined by the chosen convergence criterion. The magnifications and time delays for individual microimages are calculated analytically in our method, and the maximum  relative difference in magnification is smaller than $10^{-\textbf{5}}$ in general. 
The tests confirms that our method can identify microimages with a sufficiently high success rate and can accurately predict the position, magnification, and time delay of the identified microimages. The performance of our method generally improves when $r_l$ becomes smaller and/or $r_s$ becomes larger, which at the same time increases the computational cost. In practise, these two parameters need to be optimised in accordance with the computational resource available.

\begin{figure}
\centerline{\scalebox{1.0}
{\includegraphics[width=0.96\textwidth]{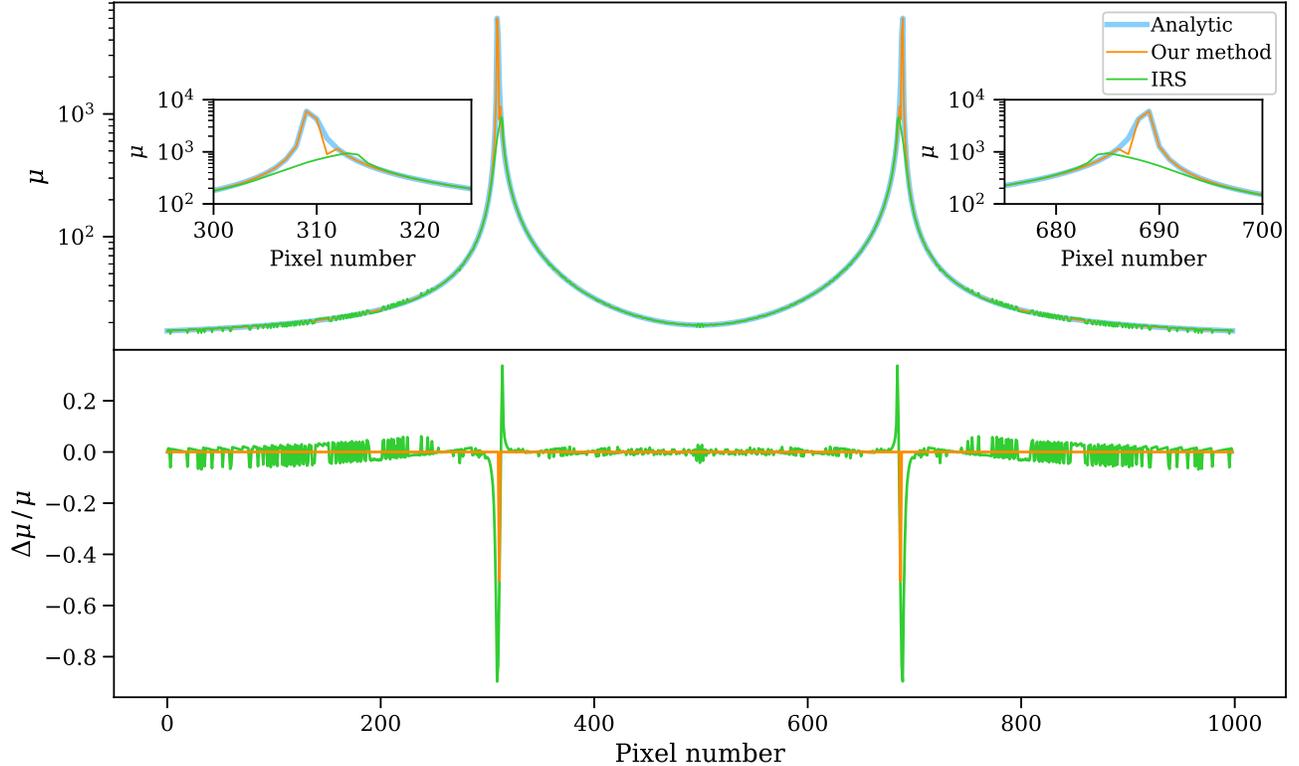}}}
\caption{\label{fig:add-fig1} \emph{(Top)}: Total magnification as a function of source position along the $y_1$ axis, i.e. horizontally along the middle of panel (a) in Figure \ref{fig:fig1}, for the single star case in section \ref{subsec:single-star}. The three curves represent predictions from the analytical solution (thick blue), our method (thin orange), and the IRS method (thin green). The two insets provide zoom-in views of the two caustic crossing regions. \emph{(Bottom)}: The relative differences in magnification of our method (thin orange) and the IRS method (thin green) with respect to the analytical predictions.
}
\end{figure}

In addition, we compute the total magnifications of microimages successfully identified by our method (with $r_l=0.005 \theta_{\rm E}$) for the 1000 source positions along $y_1$ axis and compare with the analytical predictions in Figure~\ref{fig:add-fig1}. We achieve almost perfect agreement for all source positions except for pixel 311 and pixel 687, which correspond to the two four-image cases with microimage missing in our method (last row in Table~\ref{tab:table1}). We also plot this total magnification curve generated by the IRS method in Figure~\ref{fig:add-fig1}, which shows significant deviations from the analytical predictions at almost all source positions considered. In particular, we notice that the relative differences for the IRS method show consistent fluctuations up to 7\% in the low magnification regions. Nevertheless, those discrepancies from the IRS method are understandable because the IRS method is designed to compute the total magnification for an extended area in the source plane instead of for a point and its accuracy is also limited by the resolution. 


\begin{deluxetable}{lllllll}
\tablewidth{0pt}
\tablecaption{Eight sets of parameters used for the test in Section~\ref{subsec:many-star}. $1-\kappa-\gamma$ and $1-\kappa+\gamma$ are the two eigenvalues of the Jacobian matrix. It corresponds to a minima macroimage when $1-\kappa-\gamma$ and $1-\kappa+\gamma$ are both positive and a saddle macroimage when $1-\kappa-\gamma$ and $1-\kappa+\gamma$ have opposite signs. The magnification of the macroimage $\mu$ is simply $\frac{1}{(1-\kappa-\gamma) (1-\kappa+\gamma)}$. \label{tab:paras}}
\tablehead{ 
 \colhead{} & \colhead{$\kappa$, $\gamma$, $\kappa_{\ast}$} & \colhead{$1-\kappa-\gamma$} & \colhead{$1-\kappa+\gamma$} & \colhead{$\mu$} & \colhead{$\kappa_{0}$} & \colhead{macroimage type}}
 \startdata
 Set 1 & 0.45, 0.35, 0.40 & \phantom{-1}0.2 & \phantom{1}0.9 & \phantom{-2}5.6 & \phantom{1}0.2 & minimum\\
 Set 2 & 0.45, 0.35, 0.05 & \phantom{-1}0.2 & \phantom{1}0.9 & \phantom{-2}5.6 & \phantom{1}0.2 & minimum\\
 Set 3 & 0.65, 0.55, 0.40 & \phantom{1}-0.2 & \phantom{1}0.9 & \phantom{2}-5.6 & \phantom{1}0.2 & saddle\\
 Set 4 & 0.65, 0.55, 0.05 & \phantom{1}-0.2 & \phantom{1}0.9 & \phantom{2}-5.6 & \phantom{1}0.2 & saddle\\
 Set 5 & 0.55, 0.40, 0.1 & \phantom{-}0.05 & 0.85 & \phantom{-}23.5 & 0.05 & minimum\\
 Set 6 & 0.55, 0.40, 0.0125 & \phantom{-}0.05 & 0.85 & \phantom{-}23.5 & 0.05 & minimum\\
 Set 7 & 0.60, 0.45, 0.1 & -0.05 & 0.85 & -23.5 & 0.05 & saddle\\
 Set 8 & 0.60, 0.45, 0.0125 & -0.05 & 0.85 & -23.5 & 0.05 & saddle\\\hline
 \enddata
\end{deluxetable}
\begin{figure*}
\centerline{\scalebox{1.0}
{\includegraphics[width=0.98\textwidth]{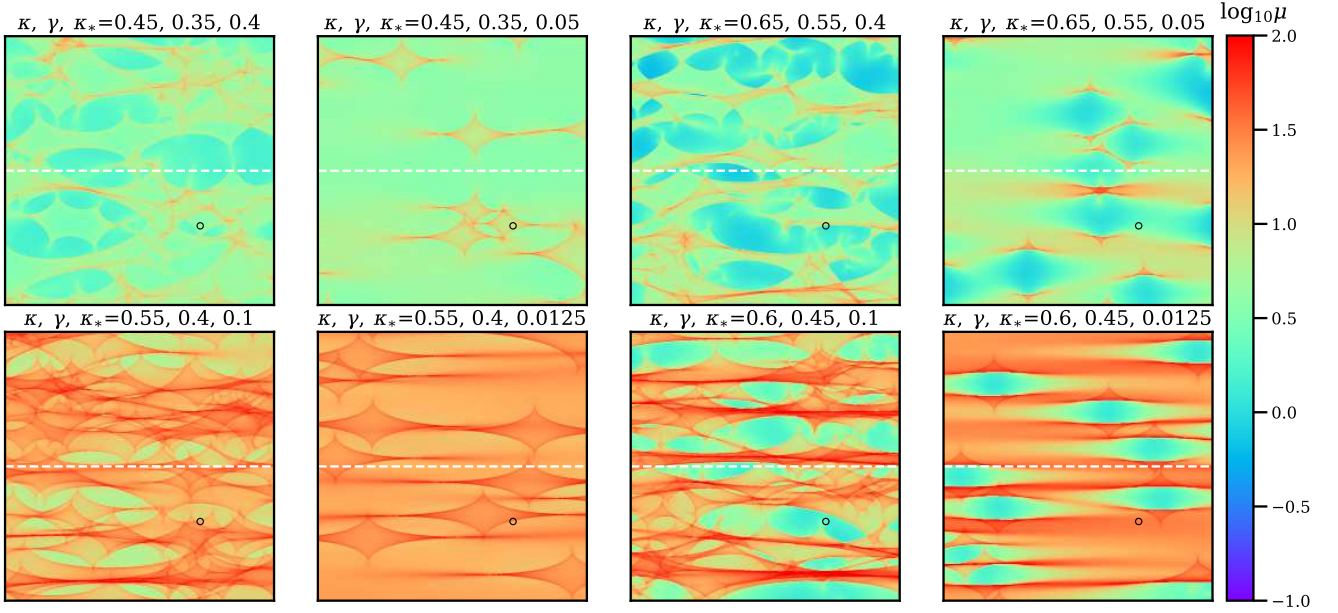}}}
\caption{\label{fig:fig3} Magnification maps generated by our method for the eight sets of parameters shown in Table \ref{tab:paras}. They are all shown on the logarithmic scale and the size of each map is $10\theta_{E} \times 10\theta_{E}$ in the source plane. The white dashed line in each map indicates the trajectory of the source corresponding to the light curve in each panel of Figure \ref{fig:fig4}. The open black circle indicates the position of the source corresponding to the light curve in each panel of Figure \ref{fig:fig5}.}
\end{figure*}
\begin{figure*}
\centerline{\scalebox{1.0}
{\includegraphics[width=0.98
\textwidth]{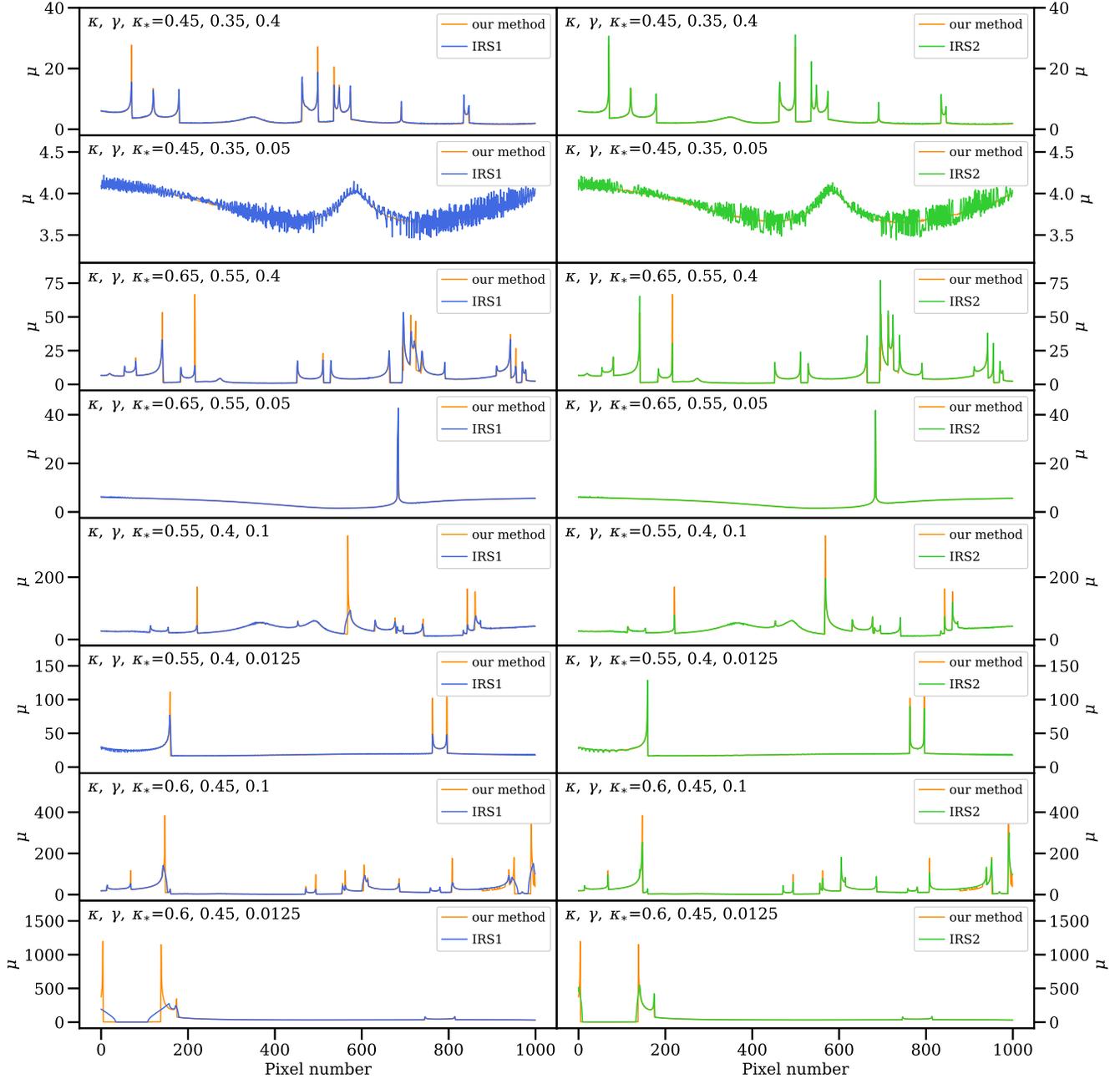}}}
\caption{\label{fig:fig4} A comparison of light curves generated by our method and the IRS method. The eight panels in the left column show a direct comparison of 1000 magnifications along the white dashed line in Figure \ref{fig:fig3}. The orange line is produced by our method and the blue line is produced by the IRS method. The eight panels in the right column show the light curves for exactly the same setups except that the source plane resolution in the IRS method is increased by a factor of five. It is clear that light curves from the IRS method agree better with light curves from our method when the resolution is increased.}  
\end{figure*}

\subsection{Many stars case}
\label{subsec:many-star}

In the case where there are many stars, the lens equation can not be solved analytically in general. We therefore choose to compare the magnification map generated by our method, i.e. by summing up the magnifications of all identified microimages, with the magnification map produced by the IRS method, which has been widely used in microlensing studies \citep[e.g.,][]{Kayser1986,1987A&A...171...49S}. We produce with our method eight magnification maps for some representative combinations of $\kappa$, $\gamma$ and $\kappa_{\ast}$, which are shown in Table \ref{tab:paras}. Specifically, $\kappa$ and $\gamma$ are chosen to represent minima macroimages in low-magnification regions in Set1 and Set2, saddle macroimages in low-magnification regions in Set3 and Set4, minima macroimages in high-magnification regions in Set5 and Set6, and saddle macroimages in high-magnification regions in Set7 and Set8, respectively. Regarding the choice of $\kappa_{\ast}$, we first introduce a characteristic star convergence $\kappa_0$ at which the average area occupied by each star is comparable to the size of the critical curve of each star. Suppose there are $N$ stars (with mass $M$) in a region under consideration with area $A$. We can define the average area occupied by each star as $A/N$ and the associated radius as $R_{\ast} = \sqrt{A/(\pi \, N)}$. Considering $\kappa_{\ast} = N \, M/(\Sigma_0 \, A)$, one can show that $R_{\ast} = \theta_{\rm E}/\sqrt{\kappa_{\ast}}$. On the other hand, the size of the critical curve of each star is approximately $\theta_{\rm E} / \sqrt{1- \kappa - \gamma}$ \citep[e.g.,][]{2018PhRvD..97b3518O}. The characteristic convergence is simply $\kappa_0 = 1 - \kappa - \gamma$. Few overlaps between the critical curves of the stars are expected when $\kappa_{\ast} < \kappa_0$ and significant overlaps are expected when $\kappa_{\ast} > \kappa_0$. We therefore choose $\kappa_{\ast} = 2 \kappa_0$ in Set1, Set3, Set5, and Set7, and $\kappa_{\ast} = \kappa_0/4$ in Set2, Set4, Set6, and Set8. The corresponding magnification maps are shown in Figure \ref{fig:fig3}. The magnification maps produced by the IRS method are not shown because the differences on those maps are very subtle. 

For a better comparison, we examine the variations of the magnification as the source moves along a given trajectory in the source plane, which is commonly referred to as  ``light curve" in quasar microlensing.  The white dashed line in Figure \ref{fig:fig3} represents the trajectory we choose, and the corresponding light curves inferred from our method and the IRS method are shown in Figure \ref{fig:fig4}. In this test, the source is assumed to be an idealised point source. For the IRS method, we consider two different source plane resolutions. In the left column of Figure \ref{fig:fig4}, the source plane resolution of the IRS method is the same as that of our method, while the resolution of the IRS method in the right column is increased by a factor of five. Overall speaking, the light curves from the two methods are in good agreement. Some discrepancies between our method and the IRS method are seen, especially near the high magnification regions. Those discrepancies become smaller when the resolution of the IRS method is increased, suggesting that our method generally produces more accurate light curves than the IRS method at the same source plane resolution. In addition, the IRS light curve for $\kappa, \gamma, \kappa_{\ast} = 0.45, 0.35, 0.05$, which corresponds to a trajectory with moderate magnifications, show consistent fluctuations again, while the light curve produced by our method for the same set of parameters is smooth. It suggests that the fluctuations are caused by the pixelation in the IRS method, which becomes particularly significant in relatively low magnification regions. With our method, the run time for the eight magnification maps ranges from several minutes to more than one hour. The corresponding run time for the high-resolution IRS method (i.e. IRS2) is from several seconds to more than one hour. Nevertheless, such a run time comparison is not fair because our method computes much more quantities.\\

\section{Microlensing of Short-duration Transients}
\label{sec:discussion}
Now we discuss a special type of microlensing phenomenon --- microlensing of short-duration transient sources such as FRBs. Since the typical duration of FRBs is on the order of millisecond, the relative motion between the lens and source within this time window is negligible. As a result, microlensing of short-duration transient sources is a ``stationary'' process that corresponds to the appearance of individual microimages with time delays. Again, those microimages can not be resolved spatially by current telescopes given the expected $\mu$as separations. However, it is feasible to temporally resolve the microimages as the time resolution of current telescopes have already reached $~\sim100 \mu$s \citep[e.g.,][]{2018MNRAS.478.1209F, 2020MNRAS.492.4752V}.\\

\begin{figure*}
\centerline{\scalebox{1.0}
{\includegraphics[width=0.98\textwidth]{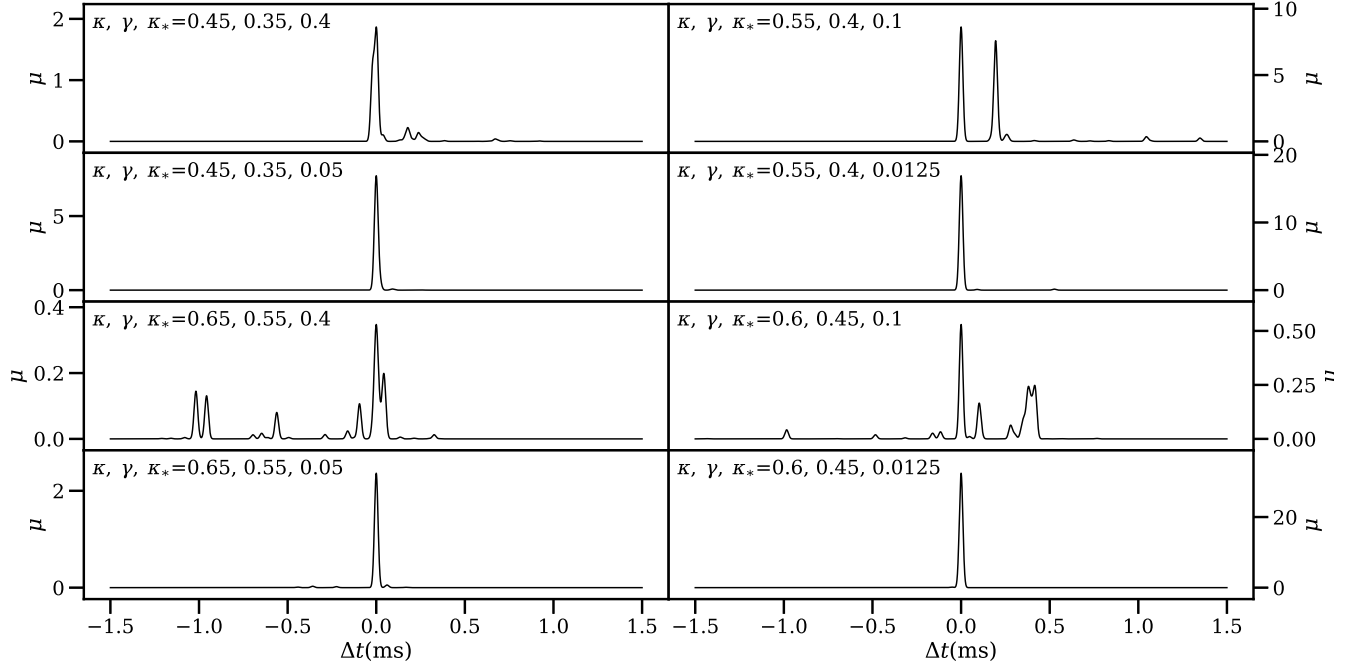}}}
\caption{\label{fig:fig5}Expected light-curves when a FRB is located at the open black circle in the eight magnification maps of Figure \ref{fig:fig3}.}
\end{figure*}

Much different from the ``light curve" (due to the flux change of the back ground source as it moves) commonly mentioned in quasar microlensing, a transient source at some position is expected to generate a microlensed light curve due to the time delays among the microimages and the magnification of each. In this case, the light curve is a superposition of pulses from individual microimages. As an example, we consider a pulse signal at a given source location (indicated by the open black circle in Figure \ref{fig:fig3}) lensed by the eight lensing fields shown in Figure \ref{fig:fig3}. In this example, the source is put at redshift $z_{s}$=0.12, which is the redshift of FRB170827 \citep{2018MNRAS.478.1209F}, and the deflector plane is set at half of the source redshift, i.e. $z_{l}$=0.06. We use our algorithm to identify microimages for each lensing field and compute their magnifications and time delays. We then assume the intrinsic pulse signal as a Gaussian with a width $\sigma$=10$\mu$s and convolve it with the magnification-time delay curves for each lensing field to obtain the light curves, which are shown in Figure \ref{fig:fig5}. The light curves are found to exhibit a wide range of structures, from a single primary peak to double peaks to complex multiple peaks. The relative time delays in these eight examples can reach 0.1--1 ms. 

Interestingly, we notice that the light curve for one lensing field ($\kappa, \gamma, \kappa_{\ast} = 0.6, 0.45, 0.1$) appears qualitatively similar to the observed light curve of FRB170827 in terms of the number of peaks and their relative flux ratios and time delays, although \citet{2018MNRAS.478.1209F} interpreted the observed temporal structures of FRB170827 as a result of a scattering screen outside the Milky Way. Nevertheless, as the number of FRBs is going to soar in upcoming surveys, microlensing of FRBs is expected to be discovered in the near future. 

In principle, the microlensing light curves encode information on $\kappa$, $\gamma$, $\kappa_{\ast}$, and the stellar mass function of the lensing galaxies at the positions of macroimages. The inference is usually under-constrained in practise given the large amount of nuisance parameters (such as the locations of the stars). Nevertheless, studies have suggested that constraints on the fractions of stars and even planets in lensing galaxies can still be obtained by analysing the observed microlensing light curves of quasars \citep[e.g.,][]{2004ApJ...605...58K, 2008ApJ...689..755M, 2011ApJ...731...71B, 2018ApJ...853L..27D}. Similar to microlensing of quasars, microlensing of FRBs can also be used to constrain those properties in distant galaxies. A major advantage of using FRBs is that the number of FRBs will keep increasing and exceed the number of quasars in a relatively short time. The total number of quasars across the full sky is on the order of $10^7-10^8$ in the LSST era \citep[e.g.,][]{2010MNRAS.405.2579O, Ivezic17}, while the Square Kilometre Array (SKA) is expected to detect $\sim 10^7$ FRBs every year \citep{Hashimoto20}. Unlike quasar microlensing where the light curves usually span a period of years during which the relative motion between the lens and source needs to be taken into account, microlensing of the short-duration FRBs can be considered as a ``stationary'' process. On the one hand, FRB microlensing has two fewer free parameters, i.e. the relative lens-source transverse velocity, which has been shown to be highly correlated with the masses of the stars and the size of the source \citep[e.g.,][]{2004ApJ...605...58K}. A faster exploration of the parameter space is also expected for FRB microlensing events. On the other hand, FRB microlensing does not probe as much the lensing field as quasar microlensing. It is therefore unclear how the constraining power is compared to quasar microlensing. We defer such an analysis to a future paper. It is worth pointing out that some FRBs are found to emit more than one bursts (i.e. repeating FRBs \citep{2016Natur.531..202S,2019Natur.566..235C,2019ApJ...885L..24C,2020ApJ...891L...6F}, and microlensing of those repeating FRBs should provide tighter constraints on the lensing field. Lastly, we note that microlensing of other types of transients such as gamma ray bursts \citep{2000MNRAS.319.1163W}, can be utilised and analysed in the same fashion as long as the duration of the transient event is sufficiently short that the change in the lens-source alignment is negligible. 
\section{Conclusions}
\label{sec:conclusion}


In this paper, we study a special type of microlensing event where the source is a short-duration transient such as FRB. Different from conventional microlensing studies that are based on light curves induced by the relative motion between the lens and source, microlensing of FRBs produces another type of light curve, which is the superposition of bursts from individual microimages. To identify individual microimages and infer their time delays and magnifications, we implement an algorithm based on two-dimensional Newton iteration method. The performance of our method mainly depends on two parameters, $r_l$ and $r_s$. In practise, they need to be chosen in accordance with the computational resource available. We note that our method is different from the IRS method commonly used in quasar microlensing studies, which can not resolve individual microimages but only provides an estimation of the total magnification. 

We perform two tests to validate our method. The first test corresponds to the case of a single point-mass lens embedded in a field of convergence and shear, for which analytical solutions exist. We compare the results given by our method and the analytical solutions for 2000 sources evenly located along the two coordinate axes and find good agreement. Above the magnification lower limit induced by the choice of $r_s$, our method reproduces all the two-image cases. For $r_{l}=0.02\theta_{E}$, our method reproduces 93\% of the four-image cases. The success rate increases to 97\% when $r_l=0.01$ and 99\% when $r_l=0.005$ (Table~\ref{tab:table1}). For the successfully identified microimages, our method can reproduce the image positions with an accuracy of better than 1--5$\times 10^{-6} \theta_{\rm E}$, and the relative difference in the predicted magnification is no larger than $10^{-5}$ in general (Figure~\ref{fig:fig2}). In addition, we find almost perfect agreement on the total magnifications between our method and the analytical solution. The total magnifications from the IRS method show noticeable deviations from the analytical solution in both low- and high-magnification regions. 

The second test corresponds to a more general situation with multiple point-mass lenses, for which analytical solutions are not available. We test our method by comparing magnification maps and light curves generated by our method with those generated with the IRS method for eight example lensing fields (Figure~\ref{fig:fig3}). We find good agreement in general despite the completely different strategies of generating the light curves in the two methods (Figure~\ref{fig:fig4}). The agreement improves when the source plane resolution in the IRS method is increased.

We further show some model light curves for FRBs that are microlensed by the eight example lensing fields used in the second test (Figure~\ref{fig:fig5}). The light curves exhibit a wide range of structures, from a single primary peak to double peaks to complex multiple peaks. The relative time delays in these eight examples can reach 0.1--1 ms, which can already be resolved by current telescopes. Although not yet discovered, microlensing of FRBs is expected to become regular events in the upcoming SKA era when $\sim 10^{4-5}$ FRBs will be detected on a daily basis. Light curves produced by our method are therefore essential for constraining the stellar fraction and stellar mass distribution in distant galaxies with FRB microlensing.

\section{Acknowledgements}
The authors thank the anonymous referee for helpful comments and thank Xuefeng Wu, Jun Zhang and Zuhui Fan for helpful discussions and suggestions. This work is supported by the NSFC (No. 11673065, U1931210, 11273061). We acknowledge the cosmology simulation database (CSD) in the National Basic Science Data Center (NBSDC) and its funds the NBSDC-DB-10 (No. 2020000088). We acknowledge the science research grants from the China Manned Space Project with NO.CMS-CSST-2021-A13. Yiping Shu acknowledges support from the Max Planck Society and the Alexander von Humboldt Foundation in the framework of the Max Planck--Humboldt Research Award endowed by the Federal Ministry of Education and Research.

\end{document}